\documentclass[twocolumn,pre,aps,showpacs,amsmath,amssymb,superscriptaddress]{revtex4}

\usepackage{amsmath}
\usepackage{graphicx}

\newcommand{\be}{\begin{equation}}
\newcommand{\ee}{\end{equation}}
\newcommand{\bea}{\begin{eqnarray}}
\newcommand{\eea}{\end{eqnarray}} 
\newcommand{\ba}{\begin{array}}
\newcommand{\ea}{\end{array}}

\newcommand{\bb}{\bibitem}

\begin{document}

\title{\bf Massive minimal subtraction scheme and ``partial-$p$'' in anisotropic Lifshitz space(time)s}

\author{Emanuel V. Souza\footnote{e-mail:evsouza@ifi.unicamp.br}, Paulo R. S. Carvalho\footnote{e-mail:prscarvalho@ufpi.edu.br}} 
\affiliation{{\it Departamento de F\'\i sica, Universidade Federal do Piau\'\i, Campus Ministro Petr\^onio Portela, 64049-500, Teresina, 
PI Brazil}}  
\author{and Marcelo M. Leite\footnote{e-mail:mleite@df.ufpe.br}}
\affiliation{{\it Laborat\'orio de F\'\i sica Te\'orica e Computacional, Departamento de F\'\i sica,\\ Universidade Federal de Pernambuco,\\
50670-901, Recife, PE, Brazil}}

\vspace{0.2cm}
\begin{abstract}
{\it We introduce the ``partial-$p$'' operation in a massive Euclidean $\lambda\phi^{4}$ scalar field theory describing 
anisotropic Lifshitz critical behavior. We then develop a minimal subtraction a la $Bogoliubov-Parasyuk-Hepp-Zimmermann$ 
renormalization scheme. As an application we compute critical exponents diagrammatically using the orthogonal approximation at 
least up to two-loop order and show their equivalence with other renormalization techniques. We discuss possible applications of 
the method in other field-theoretic contexts.}
\end{abstract}

\vspace{1cm}
\pacs{64.60.ae; 64.60.Bd; 64.60.fd; 64.60.Kw}

\maketitle

\newpage
\par {\bf Introduction} - All relativistic higher derivative field theories were relegated to a minor role for a long time because of the 
presence of ghosts associated to higher time derivatives. Recently, it was realized that keeping second order time derivatives and 
permitting higher order space derivatives would have the virtue of maintaining unitarity, although it breaks the Lorentz invariance of the 
theory \cite{A1}. Models including gauge fields have also been built \cite{A2}, culminating with the Horava's proposal of a 
quantum gravity model where space and time scale anisotropically\cite{H}. It produced a tide of new developments: from modifications to 
general relativity \cite{SVW} to new paradigms in inflationary cosmology \cite{Wang}. From the field-theoretic perspective, a Wick rotation 
of the time coordinate permits us to make a direct comparison of these field theories (``anisotropy exponent'' $z=2$) with critical 
systems pertaining to the anisotropic Lifshitz universality classes \cite{HLS} ``living'' in Lifshitz spaces  \cite{CL1}.
\par  Competition is a mechanism which can induce anisotropy. For instance, 
anisotropic Lifshitz critical behaviors \cite{HLS} arise in a variety of real physical systems, from high-$T_{C}$ superconductors 
\cite{3,4,5} to magnetic materials \cite{Sh,Be}. A modified Ising model on a $d$-dimensional lattice describing them consists of 
first-neighbor ferromagnetic interactions competing with second-neighbor antiferromagnetic couplings along $m$ 
directions (the competition axes). The two inequivalent subspaces are the $(d-m)$- and $m$-dimensional subsets ($m \neq d$). 
Perturbatively, performing the Feynman integrals associated to this $m$-axial 
Lifshitz field theory exactly in an analytical manner is still a far-off task nowadays. A preliminary, crude approximation was developed to 
solve analytically higher-loop diagrams called ``the dissipative approximation'' which yielded the critical exponents beyond one-loop level 
for the first time using massless fields \cite{AL}. Another higher order perturbative semianalytical method employing massless fields was 
presented immediately afterwards \cite{SD}. While the dissipative approximation does not conserve momentum at higher 
order diagrams, the second alternative failed in its attempt to produce analytical answers to the exponents. From the
non-perturbative analysis, the renormalization group ($RG$) treatment of the first method could not give any information 
concerning exponents along the $m$-dimensional competition subspace, whereas the latter obtains scaling laws with unclear 
meaning. Both treatments fail to produce scaling laws in the isotropic case ($d=m$). Those problems were overcome through new 
arguments using massless fields , where two $RG$ 
independent transformations result in independent scaling relations for each subspace and 
lead to a complete determination of critical properties of this system. Moreover, the development 
of the ``orthogonal approximation'' in \cite{Leite1} represented the 
first solution in perturbation theory which allows the analytical determination of 
arbitrary loop order diagrams. It was shown to be entirely equivalent with 
the massive $RG$ formulation \cite{CL1}. From the renormalizability viewpoint, the minimal subtraction offers no difficulty in the massless theory. However, the massive minimal subtraction renormalization scheme for $m$-axial anisotropic Lifshitz critical behaviors poses a formidable 
obstacle: the manipulation of overlapping divergences. 
\par What version of ``partial-$p$'' \cite{tHV} operation should be defined in 
handling those divergences showing up in higher-loop contributions, e. g., of the one-particle irreducible ($1PI$) \cite{Amit} 
two-point vertex part? In this Letter we propose a new ``partial-$p$'' operation for these anisotropic spaces. As the noncompeting 
subspace is quadratic in derivatives, the time coordinate in the aforementioned quantum field theories can 
be identified with one coordinate of the subspace without competition, for example, in the limit $(d-m) \rightarrow 1$ and our 
construction here goes beyond the context of critical phenomena.
\par As an application, we build up a version of Bogoliubov-Parasyuk-Hepp-Zimmermann ($BPHZ$) method using minimal subtraction in order to 
compute critical exponents at least up to two-loop order. We employ the orthogonal approximation in the calculation of critical indices and 
find universal results in exact agreement with previous outcomes from massless \cite{Leite1} and massive field settings \cite{CL1} using 
normalization conditions. We conclude with a few comments on the utilization of the present method in other quantum field-theoretic models.
\par{\bf{Partial-$p$ and $BPHZ$ method}} - In massive field theories the mass sets the natural scale in the renormalization group 
approach. The anisotropic $m$-axial Lifshitz behaviors require an augmented parameter space with two mass scales characterizing the two 
subspaces involved owing to the two independent scaling transformations. The bare and renormalized fields inherit this dependence on the 
masses. The bare Euclidean Lagrangian density of the scalar field with $O(N)$ symmetry representing this critical behavior can be written as
\begin{eqnarray}\label{1}
&& \mathcal{L} = \frac{1}{2}|\bigtriangledown_{m}^{2} \phi_{0}\,|^{2} +
\frac{1}{2}|\bigtriangledown_{(d-m)} \phi_{0}\,|^{2} \nonumber\\
&& + \delta_0  \frac{1}{2}|\bigtriangledown_{m} \phi_{0}\,|^{2}
 + \frac{1}{2} \mu_{0\tau}^{2\tau}\phi_{0}^{2} + \frac{1}{4!}\lambda_{0\tau}(\phi_{0}^{2})^{2} .
\end{eqnarray}
The Lifshitz critical region is characterized by $\delta_{0}=0$ and we will use this value henceforth.  
The label $\tau$ specifies the subspace associated to the bare mass $\mu_{0\tau}$. The noncompeting $(d-m)$-dimensional corresponds 
to $\tau=1$ whereas $\tau=2$ refers to the $m$-dimensional subspace. Formally, the multiplicatively renormalized one-particle 
irreducible ($1PI$) vertex parts are the mathematical entities necessary to our discussion of the renormalization scheme. In the 
subspace $\tau=1$, we set inside the vertex parts $i)$ the external momenta along the competing subspace at zero, $ii)$ $\lambda_{02}=0$, 
$iii)$ $\mu_{02}=0$ and perform scale transformations involving only $\mu_{01}$ (and vice-versa when considering the subspace $\tau=2$; 
see below).
\par The $BPHZ$ method follows closely the conventions employed in a recent work \cite{CL2} for 
critical systems without competition. (See also the excellent description of this scheme given in the book by 
Kleinert and Schulte-Frohlinde \cite{SFK}.) We perform the 
redefinitions $\phi_{0}=Z_{\phi(\tau)}^{\frac{1}{2}}\phi$, 
$\mu_{0\tau}^{2\tau}=m_{\tau}^{2\tau}\frac{Z_{m_{\tau}^{2\tau}}}{Z_{\phi(\tau)}}$ and 
$\lambda_{0\tau}=\frac{Z_{u_{\tau}}}{Z_{\phi(\tau)}^{2}} \mu_{\tau}^{\tau \epsilon_{L}} u_{\tau}$, where $\epsilon_{L}= 4+\frac{m}{2}-d$ 
is the expansion parameter and $u_{\tau}$ are the dimensionless renormalized coupling constants. Those redefinitions make the original 
bare Lagrangian to be extended with additional extra terms, the counterterms, which generate extra Feynman diagrams order 
by order in perturbation theory.
\par The redefinitions permit the bare Lagrangian density to be rewritten in terms of renormalized amounts, namely 
\begin{eqnarray}\label{2}
&& \mathcal{L} =  \frac{1}{2}Z_{\phi(\tau)}|\bigtriangledown_{m}^{2} \phi \,|^{2} +
\frac{1}{2}Z_{\phi(\tau)}|\bigtriangledown_{(d-m)} \phi \,|^{2} \nonumber\\
&& + \frac{1}{2} m_{\tau}^{2\tau}Z_{m_{\tau}^{2\tau}}\phi^{2} + \frac{1}{4!}\mu_{\tau}^{\tau\epsilon}u_{\tau} 
Z_{u_{\tau}}(\phi^{2})^{2} ,
\end{eqnarray}
whose coefficients are $\epsilon_{L}$-dependent. The two subspaces in momentum space can be labeled collectively in the form 
$p_{\tau}= q \delta_{\tau 1} + k\delta_{\tau 2}$, where $\vec{q}=(q_{1},..., q_{d-m})$ and $\vec{k}=(k_{1},...,k_{m})$. If the external 
momenta is denoted by $P_{\tau}$, the counterterms generate additional diagrams with Feyman rules in momentum space, namely 
\begin{subequations}
\begin{eqnarray}\label{3}
&& \parbox{12mm}{\includegraphics[scale=1.0]{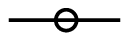}}  =  P_{\tau}^{2\tau}\delta_{\phi(\tau)}, \label{3a}\\
&& \parbox{15mm}{\includegraphics[scale=1.0]{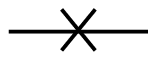}}   =  m_{\tau}^{2\tau}\delta_{m_{\tau}^{2\tau}}, \label{3b}\\
&& \parbox{8mm}{\includegraphics[scale=1.0]{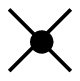}}  =  \mu_{\tau}^{\tau \epsilon}u_{\tau} \delta_{u_{\tau}}. \label{3c}
\end{eqnarray}
\end{subequations}
Note that the mass scales $\mu_{\tau}$ 
are arbitrary, $m_{\tau}$ are the renormalized masses and $Z_{\phi(\tau)}= 1 + \delta_{\phi(\tau)}, 
Z_{m_{\tau}^{2\tau}}= 1 + \delta_{m_{\tau}^{2\tau}}$ and $Z_{u_{\tau}}= 1 + \delta_{u_{\tau}}$ are the 
renormalization functions. The quantities $\delta_{\phi(\tau)}, \delta_{m_{\tau}^{2\tau}}$ and 
$\delta_{u_{\tau}}$ are the counterterms which are added at each diagram in arbitrary 
loop level in order to cancel the singular contributions of the 
primitively divergent bare vertex parts. At the loop order desired, they can be expanded in powers of the dimensionless 
renormalized coupling constants as $\delta_{\phi(\tau)}= \delta_{\phi(\tau)}^{(1)} u_{\tau} 
+ \delta_{\phi(\tau)}^{(2)} u_{\tau}^{2} + \delta_{\phi(\tau)}^{(3)} u_{\tau}^{3}$, 
$\delta_{m_{\tau}^{2\tau}} = \delta_{m_{\tau}^{2\tau}}^{(1)} u_{\tau} + \delta_{m_{\tau}^{2\tau}}^{(2)} u_{\tau}^{2}$, and 
$\delta_{u_{\tau}} = \delta_{u_{\tau}}^{(1)} u_{\tau} + \delta_{u_{\tau}}^{(2)} u_{\tau}^{2}$.    
\par Overlapping divergences can be handled by utilizing the ``partial-$p$'' operation \cite{tHV} in quadratic field theories. In 
our case, the complication is that the free propagator in momentum space is given by $\frac{1}{q^{2} + (k^{2})^{2} + m_{\tau}^{2\tau}}$. 
What saves us from that situation is the proposal of the following anisotropic version of the partial-$p$ operation:
\begin{equation}
\frac{1}{(d-m/2)}\left(\sum_{r=1}^{d-m}\frac{\partial q^{r}}{\partial q^{r}} + \sum_{s=1}^{m}\frac{1}{2}\frac{\partial k^{s}}{\partial k^{s}}\right) =1.
\end{equation}
\par And now let us compute the set of divergent integrals required in the evaluation of critical exponents at least up to two-loop order 
within this method. The diagrams required involve the one-particle irreducible ($1PI$) vertex functions $\Gamma_{\tau}^{(2)}$ and 
$\Gamma_{\tau}^{(4)}$. We actually need the singular parts of these integrals. (Denote the extraction of the singular 
parts of an arbitrary singular integral $I$ by $(I)_{S}$ \cite{CL2,Comment,BPH,Z} although we can actually neglect the subscript, 
provided this does not cause any confusion to the reader). The graphs of $\Gamma_{\tau}^{(2)}$ up to three-loop 
level can be better understood if we divide them in two main categories: the tadpole ones and the diagrams which depend on the external 
momenta (together with 
their counterterms). For the purposes of computing 
$Z_{\phi(\tau)}$ up to three-loop level, we shall need only the one- and two-loop tadpole graphs associated, respectively, with the 
integrals $I_{T1}=\int d^{d-m}qd^mk\frac{1}{q^2+(k^2)^2+m_{\tau}^{2\tau}}$ and 
$I_{T2}=\int d^{d-m}q_1d^{d-m}q_2d^mk_1d^mk_2\frac{1}{[q_1^2+(k_1^2)^2+m_{\tau}^{2\tau}]^2 [q_2^2 + (k_2^2)^2 + m_{\tau}^{2\tau}]}$. They can 
be trivially computed using the partial-$p$ using standard formulas from dimensional regularization (see, e.g., Ref. \cite{Leite1}). The corresponding diagrams can be solved using the $\epsilon_{L}$ parameter as
\begin{eqnarray}
\parbox{10mm}{\includegraphics[scale=1.0]{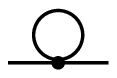}} &=& \frac{(N+2)}{3}I_{T1}= -\frac{(N+2)}{3} \frac{m_{\tau}^{2\tau}}{\epsilon_{L}} \Biggl[1+ \nonumber\\
&& \left([i_{2}]_{m} -\frac{1}{2}\right)\epsilon_{L} - 
\frac{\epsilon_{L}}{2}ln\left(m_{\tau}^{2\tau} \right) \Biggr], \label{3}\\
\parbox{10mm}{\includegraphics[scale=1.0]{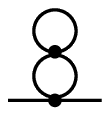}} \quad &=& \left(\frac{N+2}{3} \right)^{2} I_{T2}= -\left(\frac{N+2}{3} \right)^{2} \frac{m_{\tau}^{2\tau}}{\epsilon_{L}^2} \Biggl[1 + \nonumber\\
&& \left(2[i_{2}]_{m} -\frac{3}{2}\right)\epsilon_{L} - \epsilon_{L} ln\left(m_{\tau}^{2\tau} \right) \Biggr] \label{4},
\end{eqnarray}
where $[i_{2}]_{m}= 1+ \frac{1}{2}\Bigl[\psi(1)-\psi(2-\frac{m}{4})\Bigr]$.
\par The nontrivial higher-loop graphs of the two-point function involve explicitly the external momenta and present overlapping divergences. The 
two-loop integral 
\begin{eqnarray}
&& I_{3}= \int \frac{d^{d-m}q_{1} d^{d-m}q_{2} d^{m}k_{1} d^{m}k_{2}} {\left[q_{1}^2 + (k_{1}^2)^2 +m_{\tau}^{2\tau}\right]} 
\frac{1}{\left[q_{2}^2 + (k_{2}^2)^2 + m_{\tau}^{2\tau}\right]} \nonumber\\
&& \quad \times \frac{1}{\Bigl[(q_{1} +q_{2} + p)^2 + [(k_{1} + k_{2} + K')^{2}]^{2} +m_{\tau}^{2\tau} \Bigr]},
\end{eqnarray}
is related to the ``sunset'' diagram through $\parbox{10mm}{\includegraphics[scale=1.0]{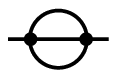}}= \left(\frac{N+2}{3} \right)I_{3}$. 
We benefit ourselves from the partial-$p$ operation by applying it twice, just as we do in conventional quadratic field theories, on the 
integrand along with the orthogonal approximation. Then, we separate the polynomials in mass and momentum 
(beside logarithms involving both). One encounters that the solution reads
\begin{eqnarray}
&& \parbox{10mm}{\includegraphics[scale=1.0]{fig6MSLaniso.eps}} = -\Bigl(\frac{N+2}{3}\Bigr) \Biggl\{ \frac{3m_{\tau}^{2\tau}}{2\epsilon_{L}^2} 
\Bigl[1+\Bigl(2[i_2]_m - \frac{1}{2}\Bigr)\epsilon_{L} \nonumber\\
&& - \epsilon_{L} ln m_{\tau}^{2\tau} \Bigr] + \frac{p^2 + (K'^2)^2}{8\epsilon_{L}} \Biggl\{1+\Bigl(2[i_2]_m-\frac{3}{4}\Bigr)\epsilon_{L} \nonumber\\
&& -  2\epsilon_{L}L_{3}(p,K',m_{\tau})\Biggr\}  \Biggr\},
\end{eqnarray}
where $L_{3}(p, K',m_{\tau})=\int_{0}^{1}dxdy(1-y) ln \Bigl\{ [p^2 + (K'^2)^2]y(1-y) + m_{\tau}^{2\tau} \Bigl[(1-y) + \frac{y}{x(1-x)}\Bigr] \Bigr\}$. 
\par The three-loop graph
\begin{eqnarray}
&& \parbox{10mm}{\includegraphics[scale=1.0]{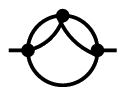}} = \left(\frac{(N+2)(N+8)}{27}\right) \int \frac{d^{d-m}q_1 d^mk_1}{[q_1^2+ (k_1^2)^2 + m_{\tau}^{2\tau}]}\nonumber \\
&&\times \quad \frac{d^{d-m}q_2 d^mk_2 d^{d-m}q_3 d^mk_3}{[q_2^2 +(k_2^2)^2 + m_{\tau}^{2\tau}][q_3^2 +(k_3^2)^2 + m_{\tau}^{2\tau}]} \times \nonumber\\
&&\frac{1}{[(q_1 + q_2 + p)^2 + [(k_1 + k_2 + K')^2]^2 + m_{\tau}^{2\tau}]} \times \nonumber \\
&& \frac{1}{\left\{(q_1 + q_3 + p)^2 + [(k_1 + k_3 + K')^2]^2 + m_{\tau}^{2\tau} \right\}} \label{intelif14},
\end{eqnarray}
can be determined by applying the partial-$p$ three times. After performing standard manipulations using the orthogonal approximation, 
at the end of the day we find the following solution:
\begin{eqnarray}
&& \parbox{10mm}{\includegraphics[scale=1.0]{fig7MSLaniso.eps}} = - \frac{(N+2)(N+8)}{27} 
\Biggl\{\frac{5m_{\tau}^{2\tau}}{3\epsilon_{L}^{3}}\Biggl[1 + \epsilon_{L}\Bigl(3[i_{2}]_{m} \nonumber\\
&& - \frac{1}{2} -\frac{3}{2}ln m_{\tau}^{2\tau}\Bigr) + O(\epsilon_{L}^{2})\Biggr] + 
\frac{[p^2 +(K'^2)^2]}{6 \epsilon_{L}^{2}} 
\Bigl\{1 + \nonumber\\
&& (3[i_2]_m-1 - 3L_{3}(p,K',m_{\tau}))\epsilon_{L} \Bigr\} \Biggr\}.
\end{eqnarray}
Note that the first term proportional to $m_{\tau}^{2\tau}$ will only contribute to $Z_{m_{\tau}^{2\tau}}$ at three-loop order, which is 
beyond our present concern here of determining $Z_{m_{\tau}^{2\tau}}$ up to two-loop level. (We can implement $m_{\tau}=0$ in the first term 
in the determination of $Z_{\phi(\tau)}$ up to three-loops and also in all diagrams $()$ presenting these polynomials in the mass, 
symbolically as $()_{m_{\tau}^{2 \tau}=0}$). This concludes the utilization of the partial $p$-operation in the 
multiplicatively renormalized vertex parts. 
\par We can expand the 2-point vertex part including the counterterms up to three-loop order in the form:
\begin{eqnarray}\label{68}
&& \Gamma_{(\tau)}^{(2)}(P_{\tau}, m_{\tau},\mu_{\tau}^{\tau \epsilon_{L}}u_{\tau})= P_{\tau}^{2\tau} + m_{\tau}^{2\tau} + u_{\tau} 
\Biggl(\frac{\mu_{\tau}^{\tau \epsilon_{L}}}{2} 
\parbox{10mm}{\includegraphics[scale=1.0]{fig4MSLaniso.eps}} \nonumber\\
&& + m_{\tau}^{2\tau}\delta_{m_{\tau}^{2\tau}}^{(1)} + P_{\tau}^{2\tau}\delta_{\phi(\tau)}^{(1)}\Biggr) 
+ u_{\tau}^{2}\Bigl( - \quad \frac{\mu_{\tau}^{2\tau \epsilon_{L}}}{4}
\parbox{10mm}{\includegraphics[scale=1.0]{fig5MSLaniso.eps}} \nonumber \\
&& - \frac{\mu_{\tau}^{2 \tau \epsilon_{L}}}{6}
\parbox{12mm}{\includegraphics[scale=1.0]{fig6MSLaniso.eps}}  - \frac{\mu_{\tau}^{\tau \epsilon_{L}}m_{\tau}^ {2\tau}\tilde{\lambda}_{m_{\tau}^ {2\tau}}}{2u_{\tau}}
\parbox{6mm}{\includegraphics[scale=1.0]{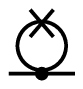}} + \frac{\mu_{\tau}^{\tau \epsilon}\tilde{\lambda}_{u_{\tau}}}{2u_{\tau}}
\parbox{10mm}{\includegraphics[scale=1.0]{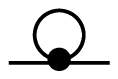}}  \nonumber\\
&& +  m_{\tau}^{2\tau}\delta_{m_{\tau}^{2\tau}}^{(2)} + P_{\tau}^{2\tau}\delta_{\phi(\tau)}^{(2)}\Bigr) 
+ u_{\tau}^{3}\Bigl(\frac{\mu_{\tau}^{3 \tau \epsilon_{L}}}{4}
\Bigl[\parbox{12mm}{\includegraphics[scale=1.0]{fig7MSLaniso.eps}}\Bigr]_{m_{\tau}^{2\tau}=0} \nonumber\\
&& - \frac{\mu_{\tau}^{2\tau \epsilon_{L}}\tilde{\lambda}_{u_{\tau}}}{3u_{\tau}}
  \Bigl[\parbox{11mm}{\includegraphics[scale=1.0]{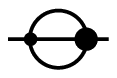}}\Bigr]_{m_{\tau}^{2\tau}=0} + P_{\tau}^{2\tau}\delta_{\phi(\tau)}^{(3)} + tadps. \Bigr), 
\end{eqnarray}
where $tadps$ mean all tadpole diagrams at three-loop order, which only contribute to $\delta_{m_{\tau}^{2\tau}}^{(3)}$ and can be safely 
neglected. Since the counterterms pick out only the singular terms of the diagrams, we conclude that 
$\delta_{m_{\tau}^{2 \tau}}^{(1)} = \frac{(N+2)}{6 \epsilon_{L}}, \delta_{\phi(\tau)}^{(1)}=0$. Let us make a pause to discuss the simplest aspect of 
the four-point vertex part in order to achieve a better comprehension of the counterterm loop diagrams. 
\par We momentarily focus on the four-point vertex function at one-loop. Its diagrammatic expansion is
\begin{eqnarray}\label{72}
&& \Gamma_{(\tau)}^{(4)}(p_{i \tau},m_{\tau},\mu_{\tau}^{\tau \epsilon} u_{\tau}) =  u_{\tau} \mu_{\tau}^{\tau \epsilon_{L}} \Bigl( 1 - u_{\tau}
\frac{\mu_{\tau}^{\tau \epsilon_{L}}}{2} \times \nonumber\\
&& \Bigl(\Bigl[\parbox{10mm}{\includegraphics[scale=1.0]{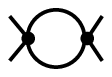}}\Bigr](p_{1 \tau}+p_{2 \tau}) + 2 \;perms. \Bigr) 
+ u_{\tau}\delta_{u_\tau}^{(1)} \Bigr),
\end{eqnarray}  
where 
\begin{eqnarray}
&& \parbox{10mm}{\includegraphics[scale=1.0]{fig11MSLaniso.eps}} (P_{\tau}) =  \frac{(N+8)}{9} \int \frac{d^{d-m}qd^{m}k}
{(q^{2}+ (k^{2})^{2}+ m_{\tau}^{2\tau})} \times \nonumber\\
&& \frac{1}{[(q+P)^{2} + ((k+K')^{2})^{2} +m_{\tau}^{2\tau}]},
\end{eqnarray}
with $P_{1}=P$ and $P_{2}=K'$. This diagram can be calculated and yields
\begin{eqnarray}
\parbox{10mm}{\includegraphics[scale=1.0]{fig11MSLaniso.eps}} (P_{\tau})&=& \frac{1}{\epsilon_{L}} \Bigl[1+ ([i_2]_m - 1 - \frac{1}{2} L(P_{\tau}))\epsilon_{L} \Bigr],
\end{eqnarray}
where $L(P_{\tau})=\int_{0}^{1}dx ln[(P^2+(K'^2)^2)x(1-x)+m_{\tau}^{2\tau})]$. Finiteness of the last vertex part at one-loop 
implies that $\delta_{u_{\tau}}^{(1)} = \frac{(N+8)}{6\epsilon_{L}}$.
\par Coming back to the two-point vertex part, the counterterm diagram 
$\parbox{6mm}{\includegraphics[scale=1.0]{fig8MSLaniso.eps}}$ can be understood as the four-point one-loop diagram computed at zero 
external momentum, with the upper coupling constant replaced by $\tilde{\lambda}_{m_{\tau}^ {2\tau}}= u_{\tau} \delta_{m_{\tau}^{2 \tau}}^{(1)}$. 
Expanding $\mu_{\tau}^{\tau \epsilon_{L}}$ in powers of $\epsilon_{L}$ the contribution of the counterterm is given by:
\begin{eqnarray}\label{71}
&& \frac{\mu_{\tau}^{\tau \epsilon_{L}} m_{\tau}^{2\tau} \tilde{\lambda}_{m_{\tau}^{2\tau}}}{2u_{\tau}}
 \parbox{6mm}{\includegraphics[scale=1.0]{fig8MSLaniso.eps}} = m_{\tau}^{2\tau}\frac{(N+2)^{2}}{36\epsilon_{L}^{2}}\Biggl[1 + ([i_{2}]_{m}-1)\epsilon_{L} \nonumber\\
&& - \frac{1}{2}\epsilon_{L} \ln\left(\frac{m_{\tau}^ {2\tau}}{\mu_{\tau}^{2\tau}}\right)\Biggr].
\end{eqnarray}
The next counterterm diagram is the product of a one-loop tadpole with a one-loop four-point insertion where the loop has shrunken to zero, 
picking out the coupling constant at $\tilde{\lambda}_{u_{\tau}}= u_{\tau}\delta_{u_{\tau}}^{(1)}$. Similar expansions in $\epsilon_{L}$ 
as performed in the previous counterterm diagram lead us to 
\begin{eqnarray}\label{74}  
&&\frac{\mu_{\tau}^{\tau\epsilon_{\tau}}\tilde{\lambda}_{u_{\tau}}}{2u_{\tau}}
\parbox{10mm}{\includegraphics[scale=1.0]{fig9MSLaniso.eps}}  = - m_{\tau}^{2} \frac{(N+2)(N+8)}{36\epsilon_{L}^{2}} 
\Bigl[1 + ([i_{2}]_{m}-\frac{1}{2}) \nonumber\\
&& \times  \epsilon_{L} - \frac{\epsilon_{L}}{2} ln\Bigl(\frac{m_{\tau}^{2\tau}}{\mu_{\tau}^{2\tau}}\Bigr)\Bigr].
\end{eqnarray}  
Combining the two-loop contribution eliminates all the terms proportional to $ln\Bigl(\frac{m_{\tau}^{2\tau}}{\mu_{\tau}^{2\tau}}\Bigr)$ at 
$O(u_{\tau}^{2})$. The divergences are cancelled provided $\delta_{m_{\tau}^{2\tau}}^{(2)}=\frac{(N+2)(N+5)}{36\epsilon_{L}^{2}} - \frac{(N+2)}{24\epsilon_{L}}$ and $\delta_{\phi(\tau)}^{(2)}= - \frac{(N+2)}{144\epsilon_{L}}$. The three-loop counterterm diagram is 
the ``sunset'' with one of the couplings replaced by the coupling constant counterterm at one-loop. The combination of the diagrams 
at this loop order in the simplified form displayed eliminates the $L_{3}(p,K',m_{\tau})$ and we can read off the value 
$\delta_{\phi(\tau)}^{(3)}= - \frac{(N+2)(N+8)}{1296\epsilon_{L}^{2}}$.
\par We are left with the computation of $\delta_{u_{\tau}}^{(2)}$. The two-loop contribution of the four-point vertex part 
is given by
\begin{eqnarray}\label{76}
&&\Gamma_{2-loop}^{(4)}(k_{i},m, \mu^{\epsilon}u) = \mu^{\epsilon}u^{3} 
\Bigl[ \frac{\mu^{2\epsilon}}{4}
         \Bigl(\Bigl[\parbox{16mm}{\includegraphics[scale=1.0]{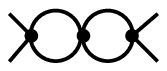}}\Bigr](k_{1}+k_{2}) \nonumber\\
&& + 2perms.\Bigr) + \frac{\mu^{2\epsilon}}{2}
   \Bigl(\Bigl[\parbox{10mm}{\includegraphics[scale=1.0]{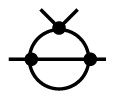}}\Bigr](k_{i}) + 5perms. \Bigr) 
+ \frac{\mu^{2\epsilon}}{2} \times \nonumber\\
&&  \Bigl(\Bigl[\parbox{11mm}{\includegraphics[scale=1.0]{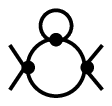}}\Bigr](k_{1}+k_{2}) + 2perms. \Bigr) 
+ \frac{\mu^{\epsilon} m^{2} \tilde{\lambda}_{m^ {2}}}{2u}
\Bigl(\Bigl[\parbox{11mm}{\includegraphics[scale=1.0]{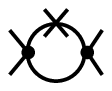}}\Bigr] \nonumber\\
&& (k_{1}+k_{2}) + 2perms. \Bigr)\Bigr)\;
- \frac{\mu^{\epsilon}\tilde{\lambda}_{u}}{u}\Bigl(\Bigl[\parbox{11mm}{\includegraphics[scale=1.0]{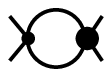}}\Bigr](k_{1}+k_{2}) \nonumber\\ 
&& + 2perms. \Bigr) + \;\delta_{u}^{(2)}\Bigr].
\end{eqnarray} 
The first diagram is the one-loop contribution to the four-point function to the square. The nontrivial two-loop diagram of the 
four-point function 
\begin{eqnarray}
&&[\parbox{10mm}{\includegraphics[scale=1.0]{fig13MSLaniso.eps}}\Bigr](P_{\tau})= \left(\frac{(N+2)(N+8)}{27}\right)
\int d^{d-m}q_1 d^{d-m}q_2 \nonumber \\
&& \frac{d^{d-m}q_3 d^mk_1 d^mk_2 d^mk_3}{[q_1^2+ (k_1^2)^2 + m_{\tau}^{2\tau}][q_2^2 +(k_2^2)^2 + m_{\tau}^{2\tau}][q_3^2 +(k_3^2)^2 + m_{\tau}^{2\tau}]} \nonumber \\
&&\quad \times\frac{1}{\left\{(q_1 + q_2 + p)^2 + [(k_1 + k_2 + K')^2]^2 + m_{\tau}^{2\tau} \right\}} \times \nonumber \\
&& \frac{1}{\left\{(q_1 + q_3 + p)^2 + [(k_1 + k_3 + K')^2]^2 + m_{\tau}^{2\tau} \right\}}, \label{intelif14}
\end{eqnarray}
can be expanded in $\epsilon_{L}$ as 
\begin{eqnarray}
&&[\parbox{10mm}{\includegraphics[scale=1.0]{fig13MSLaniso.eps}}\Bigr](P_{\tau})= \frac{(N+2)(N+8)}{54 \epsilon_{L}^2} 
\Biggl\{1+ \left(2[i_2]_m - \frac{3}{2} \right)\epsilon_{L} \nonumber\\ 
&& \quad - \epsilon_{L}L(P_{\tau})\Biggr\}.
\end{eqnarray}
The singular part of the third diagram is 
canceled by the (counterterm) fourth graph. The last counterterm diagram can be easily constructed from our previous discussion. 
Summing up all diagrams, the $L(P_{\tau})$ contributions do vanish in the singular terms. The four-point function becomes finite 
if $\delta_{u_{\tau}}^{(2)}=\frac{(N+8)^{2}}{36\epsilon_{L}^{2}} - \frac{(5N+22)}{36\epsilon_{L}}$. 
\par The Wilson functions are defined in terms of the dimensionless coupling constants $u_{\tau}$ by 
$\beta_{\tau}(u_{\tau})= -\tau \epsilon_{\tau} \frac{\partial ln[Z_{u_{\tau}} Z_{\phi(\tau)}^{-2}u_{\tau}]^{-1}}{\partial u_{\tau}}$, 
$\gamma_{\phi}(u_{\tau}) = \beta_{\tau}(u_{\tau})
\left(\frac{\partial ln Z_{\phi(\tau)}}{\partial u_{\tau}}\right)$ and $\gamma_{m_{\tau}}(u_{\tau}) = \gamma_{\phi(\tau)}(u_{\tau}) - 
\beta_{\tau}(u_{\tau})\left(\frac{\partial  ln Z_{m_{\tau}^{2\tau}}}{\partial u_{\tau}}\right)$. The fixed point is defined by 
$\beta_{\tau}(\tilde{u}_{\tau \infty})=0$ which implies $\tilde{u}_{\tau \infty}=\frac{6\epsilon_{L}}{(N+8)}[1 
+ \frac{3(3N+14)\epsilon}{(N+8)^{2}}]$.
Through the identifications $ \eta_{\tau}=\gamma_{\phi(\tau)}(\tilde{u}_{\tau \infty})$ and 
$\nu_{\tau} = (2\tau -\gamma_{m_{\tau}}(u_{\tau}))$ we obtain the exponents
\begin{subequations}
\begin{eqnarray} 
\eta_{\tau}&=& \frac{ \epsilon_{L}^{2} \tau(N + 2)}{2(N+8)^2}
\Bigl[1 + \epsilon_{L}(\frac{6(3N + 14)}{(N + 8)^{2}} - \frac{1}{4})\Bigr],\\
\nu_{\tau}&=& \frac{1}{2\tau} + \frac{(N + 2)}{4 \tau (N + 8)} \epsilon_{L} \nonumber\\
&& + \frac{1}{8\tau}\frac{(N + 2)(N^{2} + 23N + 60)} {(N + 8)^3} \epsilon_{L}^{2}.
\end{eqnarray}
\end{subequations}
Performing the identifications $\eta_{1} \equiv \eta_{L2}$ ($\eta_{2} \equiv \eta_{L4}$), $\nu_{1} \equiv \nu_{L2}$ 
($\nu_{2} \equiv \nu_{L4}$) we retrieve the expressions for these exponents already found in refs. \cite{CL1,Leite1}. 
Utilizing the scaling laws derived in \cite{Leite1} we obtain all other exponents which are identical to those 
determined before.
\par {\bf Conclusions} The proposed $p$-partial operation for $m$-axial anisotropic Lifshitz scalar field theory does circumvent 
the problem of overlapping divergences in higher-loop Feynman integrals as explicitly demonstrated herein in the computation of the 
two-point vertex part. On the other hand, the present massive minimal subtraction in the computation of critical exponents closes the 
circle and proves the complete mathematical consistency of the orthogonal approximation with a great deal of information using quite 
different renormalization schemes and in agreement with the universality hypothesis. We wish to expand the panorama of minimal 
subtraction within this massive formulation in conjunction with the orthogonal approximation by developing the appropriate version of 
the unconventional approach first introduced in Ref. \cite{CL2} for ordinary critical systems.
\par Besides, the perturbative treatment of anisotropic quantum field theories in Lifshitz spacetimes can be greatly benefited 
from the method just developed. It might prove interesting to see how the orthogonal approximation can address the higher-loop 
computations of observables in those sort of field-theoretic models. For example, at spacetime dimension $d=D+1$ the 
Ho$\check{r}$ava-Lifshitz gravity for a careful choice of parameters has a classical Weyl invariance for $z=D$. For $z=3$ corresponding 
to four-dimensional spacetime, a scalar field coupled with this gravity system in a Weyl-invariant way was shown to possess an anomaly 
computed in position space \cite{Theisen}. With the development just obtained, we could compute certain flat space $n$-point correlators 
{\it in momentum space} at arbitrary loop order in the determination of the anomaly and make a comparison with the previous result.   
This is certainly a missing part in a better understanding of fields 
propagating in such backgrounds. Our betting is that the generation of new effects could broaden up our present knowledge of the subject, 
sheding light on these issues just as the pioneer works on the description of phase transitions did on unveiling the perturbative 
structure of ordinary (quadratic) quantum field theories. 
\par Finally, we can adapt the aforementioned technique to tackle the partial-$p$ operations in generic anisotropic competing systems 
of the Lifshitz type \cite{Leite2} where arbitrary even momentum powers are present in the free propagator. This problem is connected 
with field theories in anisotropic spacetimes with (even) arbitrary anisotropy exponent $z$. We believe it can be 
formulated similarly as discussed in the present Letter. An extended version of the present work will be presented elsewhere.
\par{\bf Acknowledgements} EVS would like to thank CAPES (Brazilian agency) for financial support. PRSC acknowledges partial support 
from FAPEPI (State of Piau\'i Foundation) and CNPq (Brazilian agency) grant number CCN-052/2010. MML was partially supported by CAPES and 
FACEPE (State of Pernambuco Foundation).

\end{document}